\documentclass[aps,prl,twocolumn,superscriptaddress]{revtex4-1}
\usepackage{amsmath,bbm,amssymb}
\usepackage{xcolor,graphics,graphicx}
\usepackage[caption=false]{subfig}
\usepackage[export]{adjustbox}

\newcommand{\be}{\begin{equation}}
\newcommand{\ee}{\end{equation}}
\newcommand{\bea}{\begin{align}}
\newcommand{\eea}{\end{eqnarray}}
\newcommand{\id}{\mathbbm{1}}

\newcommand{\sig}{{\boldsymbol{\sigma}}}

\newcommand{\rin}[1]{\hat{\boldsymbol{r}}_{in,#1}}
\newcommand{\ro}[1]{\hat{\boldsymbol{r}}_{out,#1}}
\newcommand{\re}[1]{\hat{\boldsymbol{r}}_{e,#1}}
\newcommand{\rs}{\hat{\boldsymbol{r}}}
\newcommand{\bmat}{\begin{pmatrix}}
\newcommand{\emat}{\end{pmatrix}}
\newcommand{\ranc}[1]{\hat{\boldsymbol{r}}_{anc, #1}}
\newcommand{\x}{\times}
\newcommand{\ve}{\boldsymbol{v}}

\begin{document}


\title{Ultimate squeezing through coherent quantum feedback:\\ A fair comparison with measurement-based schemes}


\author{Alfred Harwood}
\author{Alessio Serafini}
\affiliation{Department of Physics \& Astronomy, University College London, Gower Street, WC1E 6BT, London, United Kingdom}


\date{\today}

\begin{abstract}

We develop a general framework to describe interferometric coherent feedback loops and prove that, under any such scheme, the steady-state squeezing of a bosonic mode subject to a rotating wave coupling with a white noise environment and to any quadratic Hamiltonian must abide by a 
noise-dependent bound that reduces to the 3dB limit at zero temperature. 
Such a finding is contrasted, at fixed dynamical parameters, with the performance of homodyne continuous monitoring of the output modes. 
The latter allows one to beat coherent feedback and the 3dB limit under certain dynamical conditions, which will be determined exactly.

\end{abstract}


\maketitle

\noindent{\em Introduction --}
Feedback is one of the main avenues to exert and refine control on physical systems. In quantum mechanics, feedback control may be applied in two, radically different, fashions: as measurement-based feedback \cite{wisebook}, where measurements are used to purify the system and to inform operations on it, or as coherent feedback \cite{lloyd00}, where only deterministic manipulations of subsystems coupled to the system of interest (part of the latter's environment) are performed. While in measurement-based feedback the quantum information is turned into classical information by the act of measuring, in coherent feedback the information stays 
quantum at all stages of the control loop.

It might be argued that, since the manipulations involved are deterministic, coherent feedback loops should not be considered as feedback control at all, but rather as a class of open-loop control strategies where only certain auxiliary degrees of freedom are accessible. However, quantum optics allows us to disregard such a terminological dispute 
(although the discussion of this issue in \cite{jacobs_coherent_2014} is worth mentioning), through the adoption of the input-output formalism, which is tailored to describe the interaction of a countable set of localised modes (e.g., a set of cavity modes) with a neighbouring field's continuum (the electromagnetic field outside a cavity). We shall then, as customary in the context of quantum optics, define a coherent feedback loop as one where a set of output modes, interacting with a system at an input-output interface, 
may be manipulated through quantum CP-maps and then fed back into a system as input modes at another input-output interface. This approach is similar to that used in the established field of `cascaded quantum systems', where the output of one system is used as the input of another \cite{gardiner_driving_1993,carmichael_quantum_1993,giovannetti_master_2012}, and of the related ``all optical'' feedback \cite{wiseman94}, the difference being that here the output is fed back into the system whence it came.

Notice that the input-output paradigm finds successful and broad application to a number of quantum set-ups
where a high degree of coherent control is achievable, ranging from purely optical set-ups to optomechanics \cite{aspelmeyer_cavity_2014}, nanoelectromechanics, atomic ensembles,  and cavity QED waveguides \cite{lalumiere_input-output_2013}, to mention but a few. Given the impressive recent advances in the realisation of quantum technologies, connectivities 
(especially via fibres and waveguides) are quickly nearing a point where quantum control loops will be feasible and pivotal in harnessing quantum resources, such as quantum coherence and entanglement. Indeed, coherent control loops have been demonstrated in optical \cite{iida_experimental_2012} and solid-state \cite{hirose_coherent_2016} systems, 
whilst measurement-based feedback has been by now applied to a variety of systems, with the aim of performing quantum operations, of enhancing cooling routines \cite{schafermeier_quantum_2016} or of entangling quantum systems \cite{riste_deterministic_2013}. 
It is therefore paramount to understand the ultimate limits of feedback strategies as well as which class of loops, coherent or measurement-based, is advantageous to perform a given task or optimise a given figure of merit. 

After the seminal study \cite{wiseman94} -- which, at variance with the present enquiry, does not deal with a squeezing Hamiltonian acting on the system, but rather with more general forms of coupling between system cavity and the feedback loop, 
such a theoretical comparison has been addressed only for specific tasks in finite-dimensional scenarios \cite{jacobs_coherent_2014,balouchi17}, or for the realisation of protocols involving out-of-loop degrees of freedom \cite{yamamoto_coherent_2014} (i.e., concerning the relationship between input and output degrees of freedom). 
Treatment \cite{yamamoto_coherent_2014}, in particular, adopts a framework that is wholly analogous to ours, and establishes a few remarkable impossibilities for measurement-based feedback. In our study, the input-output formalism, compounded with general linear operations, will yield a framework for a fair comparison between the two classes of control schemes, which may be contrasted at given connectivities and other technical and environmental parameters (such as detection efficiencies and temperature). 

Note that the set of operations encompassed by coherent and measurement-based feedback differ, since  
the stochastic dynamics originating from measurements cannot be reduced to deterministic operations, 
which is the essence of the so called ``measurement problem'' of quantum mechanics. 
Nevertheless, coherent feedback has been proven superior in a number of tasks and contexts, so that our ability to demonstrate situations where measurement-based schemes do in principle prove superior is all the more striking and consequential. Measurement-based strategies, it turns out, prove particularly effective 
in stabilising in-loop figures of merit (i.e., quantities pertaining to the localised modes).
 
Specifically, in this paper we shall consider a single bosonic mode 
coupled to a white noise continuum at finite temperature through the input-output formalism. 
We shall assume the mode to be subject to a squeezing Hamiltonian and, as a significant case study, shall adopt the optimisation of steady-state squeezing as our figure of merit. First, as proof of principle, we shall present a simple coherent feedback loop using a single feedback mode subject to losses, show that it can enhance the achievable squeezing, and contrast it with what is achievable through (feasible) homodyne measurements of the output field. 
Then, we consider the most general possible coherent feedback setup, letting an arbitrary number of output modes at one interface undergo the most general deterministic Gaussian CP-map not involving any source of squeezing (i.e., the most general open, passive optical transformation, corresponding in practice to leakage, beam splitters and phase shifters), before being fed back into the system. This will prove that the simple scheme we considered 
is indeed optimal, and that our comparison is therefore conclusive.\smallskip

\noindent{\em Continuous variable systems -- }
A system of $n$ bosonic modes can be described as a vector of operators $\hat{\boldsymbol{r}} = (\hat{x}_1,\hat{p}_1...\hat{x}_n,\hat{p}_n)^{\sf T}$. These obey canonical commutation relations (CCR) $[\hat{x}_i, \hat{p}_j] = i \delta_{ij}\hat{\id} $ where we have set $\hbar=1$. The CCR for multiple modes can be described using the symmetrised version of the commutator $ [\hat{\boldsymbol{r}}, \hat{\boldsymbol{r}}^{\sf T}] = \hat{\boldsymbol{r}} \hat{\boldsymbol{r}}^{\sf T} - (\hat{\boldsymbol{r}}\hat{\boldsymbol{r}}^{\sf T})^{\sf T} =i\Omega_n$ where $\Omega_n$ is a $2 n \times 2 n$ matrix known as the symplectic form:
$\Omega_n = \bigoplus_{j=1}^n \Omega_1$, with 
 $\Omega_1 = \begin{pmatrix} 0 & 1 \\
                              -1 & 0\end{pmatrix}$.
In the rest of this paper, we will omit the subscript from $\Omega$, letting the context specify the appropriate dimension. For a quantum state $\hat{\rho}$, the expectation value of the observable $\hat{x}$ is given by $\langle \hat{x} \rangle =  \text{Tr}[\hat{\rho} \hat{x}]$. Using  vector notation, this can be generalised to give the first and second statistical moments of a state:
\be \label{moments}
    \boldsymbol{\Bar{r}} = \text{Tr}[\hat{\rho} \hat{\boldsymbol{r}}] \quad \quad
    \boldsymbol{\sigma} = \text{Tr}[\{(\hat{\boldsymbol{r}} - \Bar{\boldsymbol{r}}), (\hat{\boldsymbol{r}} - \Bar{\boldsymbol{r}})^{\sf T}\}\hat{\rho}] \, .
\ee
The above definition leads to a real, symmetric covariance matrix $\sig$.

The steady-states we will focus on are 
Gaussian states, which may be defined as the ground and thermal states of quadratic Hamiltonians. Such states are fully characterised by first and second statistical moments, as defined above. 
Unitary operations which map Gaussian states into Gaussian states are those generated by a quadratic Hamiltonians. The effect of such operations on the vector of operators is a symplectic transformation $\hat{\boldsymbol{r}} \rightarrow S \hat{\boldsymbol{r}}$ where $S$ is a $2n \times 2n$ real matrix which satisfies $S\Omega S^{\sf T} = \Omega$. The corresponding effect on the covariance matrix of the system is the transformation $ \boldsymbol{\sigma} \xrightarrow{} S \boldsymbol{\sigma}S^{\sf T}$.  In this study, we will make use of so-called `passive' transformations, which do not add any energy to the system and therefore do not perform any squeezing. Passive transformations must satisfy the extra constraint that $S$ is orthogonal, ie. $SS^{\sf T} = \id$. \smallskip


\noindent{\em The Input-Output Formalism --}
The input-output formalism is a method for dealing with the evolution of systems coupled to a noisy environment, consisting of a continuum of modes (e.g., the free electromagnetic field). The interaction of the system with such an environment can be modelled as a series of instantaneous interactions with different modes at different times \cite{gardiner_input_1985}. 
The incoming mode which interacts with the system at time $t$ is known as the input mode and is labelled $\hat{x}_{in}(t)$, The mode scattered at time $t$ is labelled $\hat{x}_{out}(t)$ and is known as the output mode. The input modes satisfy the 
continuous CCR: $[\hat{\boldsymbol{r}}_{in}(t), \hat{\boldsymbol{r}}_{in}^{\sf T}(t')] = i\Omega \delta(t-t')$ where $\hat{\boldsymbol{r}}_{in}(t) = ( \hat{x}_{in,1}, \hat{p}_{in,1}...\hat{x}_{in,m}, \hat{p}_{in,m})^{\sf T}$. The coupling of the system to the input fields is given by a Hamiltonian $\hat{H}_C$:
\be
    \hat{H}_C = \frac{1}{2}\hat{\boldsymbol{r}}_{SB}^{\sf T}H_C\hat{\boldsymbol{r}}_{SB} = \frac{1}{2}\hat{\boldsymbol{r}}_{SB}^{\sf T} \begin{pmatrix} 0 & C \\ C^{\sf T} & 0 \end{pmatrix} \hat{\boldsymbol{r}}_{SB} \, ,
\ee
where $ \hat{\boldsymbol{r}}_{SB} = (\hat{x}_1, \hat{p}_1, ...\hat{x}_n, \hat{p}_n, \hat{x}_{in,1}, \hat{p}_{in,1}...\hat{x}_{in,m}, \hat{p}_{in,m})^{\sf T}$ is the total vector of the $n$ system modes and $m$ bath modes. The square matrix $H_C$ is the coupling Hamiltonian matrix and $C$ is a $2n \times 2m$ matrix known as the coupling matrix. The Heisenberg evolution of the system operators is given by a stochastic differential equation known as the quantum Langevin equation \cite{serafini_quantum_2017}:
\be \label{Langevin}
    {\rm d} \hat{\boldsymbol{r}}(t) = A \hat{\boldsymbol{r}}(t)\, {\rm d}t + \Omega C \hat{\boldsymbol{r}}_{in}(t)\, {\rm d}t
\ee
The matrix $A$ is known as the drift matrix of the system and is given by $A = \Omega_n H_S + \frac{1}{2}\Omega C \Omega C^{\sf T}$. The symmetric square matrix $H_S$ specifies the system Hamiltonian $\hat{H}_S = \frac{1}{2}\rs^{\sf T}H_S\rs$. The vector $\hat{\boldsymbol{r}}_{in}(t)$ is a stochastic process known as a quantum Wiener process which, in analogy with the classical Wiener process, obeys the relations 
$[\hat{\boldsymbol{r}}_{in}(t), \hat{\boldsymbol{r}}_{in}^{\sf T}(t)]({\rm d}t)^2 = i \Omega \,{\rm d}t$ and $\langle \{\hat{\boldsymbol{r}}_{in}(t), \hat{\boldsymbol{r}}_{in}^{\sf T}(t)\} \rangle({\rm d}t)^2 =  \boldsymbol{\sigma}_{in} {\rm d}t$ where $\boldsymbol{\sigma}_{in}$ is the covariance matrix of the input modes. 
This relationship implies delta correlations between bath modes interacting at different times (the well-known ``white noise'' condition)
and hence the Markovianity of the free dynamics, which we are thus assuming.

Eqs. (\ref{moments}) and (\ref{Langevin}) can be combined to obtain an equation for the evolution of the system covariance matrix:
\be \label{dde}
    \dot{\boldsymbol{\sigma}} = A \boldsymbol{\sigma} + \boldsymbol{\sigma} A^{\sf T} + D \; , 
\ee
where $D = \Omega C \boldsymbol{\sigma}_{in} C^{\sf T} \Omega^{\sf T}$ is known as the diffusion matrix. 
The condition required for Eqs.~(\ref{dde},\ref{Langevin}) to admit a steady-state (stable) solution is that the matrix $A$ must be `Hurwitz', meaning that all of the real parts of its eigenvalues are negative. If this condition is satisfied, then the steady-state solution reads
\be \label{steadystate} 
    \boldsymbol{\sigma}_{\infty} = \int_{0}^{\infty} e^{A t}De^{A^{\sf T} t}\, {\rm d}t \, .
\ee

\noindent{\em Squeezing with no control --}
Squeezing is the process of reducing the variance of one quadrature and correspondingly increasing the variance of its conjugate. Our figure of merit for this study is $\sigma_{11}$, the element of the covariance matrix corresponding to twice the variance of the $\hat{x}$-quadrature. The smaller the value of $\sigma_{11}$, the more squeezed the system is. We consider a single bosonic cavity mode, subject to the Hamiltonian $\hat{H} = \hat{H}_S + \hat{H}_C$, where $\hat{H}_S = -\chi \{\hat{x},\hat{p}\}/4$ with $\chi>0$ is the Hamiltonian which squeezes the $\hat{x}$ quadrature.  This corresponds to a Hamiltonian matrix $H_S = - \frac{\chi}{2} \sigma_x$, 
where $\sigma_x$ is the Pauli $x$-matrix. Losses in the cavity are modeled by coupling the cavity mode to an external fields through the Hamiltonian $\hat{H}_C $ that allows for the exchange of excitations:
\be \label{coupling}
    \hat{H}_C = \sqrt{\gamma} (\hat{p}\hat{x}_{in} - \hat{x}\hat{p}_{in}) \; ,
\ee 
This corresponds to a coupling matrix $C =  \sqrt{\gamma}\Omega_1^{\sf T}$ where $\gamma$ is the strength of the coupling. When this is the form of system-environment coupling, the so-called input-output boundary condition relates the system modes to the input and output as follows \cite{gardiner_input_1985}:
\be \label{boundary}
    \hat{\boldsymbol{r}}_{out}(t) =  \sqrt{\gamma}\hat{\boldsymbol{r}}(t) - \hat{\boldsymbol{r}}_{in}(t) \, .
\ee
For this example and for the remainder of this paper, we will consider the case when all input fields are in Gibbs thermal states of the free Hamiltonian, so that $\boldsymbol{\sigma}_{in} = \Bar{N}\id$, where $\Bar{N} = 2N +1$ and $N\geq0$ is the mean number of thermal excitations in the environment,
which can be promptly related to temperature and frequency through the Bose law. 
The steady-state squeezing obtained from (\ref{steadystate}) is $\sigma_{11} = \Bar{N}\gamma/(\chi + \gamma)$. The condition for stability is that $|\chi|<\gamma$, which means that, if the loss rate or squeezing parameter can be tuned, and the input fields are taken to be vacua (so $\Bar{N}=1$) the maximum steady-state squeezing that can be achieved is $\sigma_{11} =  \frac{1}{2}$. This is known in the literature as the 3dB limit, as $10\log_{10} (2) \approx 3.01$ (this is the noise, in decibels, associated with the smallest eigenvalue 
of $\sig$ in units of vacuum noise).

We note that steady-state squeezing could be improved upon if the input state were squeezed, 
but in this study we will only consider naturally occurring, non-squeezed reservoirs (as opposed to squeezed ones, which have only been 
envisaged through very demanding engineering), also in view of comparing different feedback strategies under the assumption that the only 
squeezing source is constituted by the system Hamiltonian with a certain strength $\chi$.\smallskip

\noindent{\em Homodyne Monitoring --}
In order to provide the reader with a comparison between measurement-based and coherent feedback, let us also consider 
the continuous monitoring of the output $\hat{x}$-quadrature through a homodyne detector with efficiency $\zeta$, which yields 
a relevant element of the covariance matrix given by \cite{SM}:	
\be
   \sigma_{11}^m = \frac{a+\sqrt{a^2+b}}{2\zeta}   \; , \label{moni}
\ee
for $a= [2 \Bar{N}\zeta -(1+(\Bar{N}-1)\zeta)(1+\frac{\chi}{\gamma})]$ and 
  $b = 4 \Bar{N} \zeta (1-\zeta)$.
This monitoring maximises the steady-state squeezing among all general-dyne detections  
at zero temperature, i.e., for $\Bar{N}=1$ \cite{genoni_optimal_2013}, but is beneficial at finite temperature too (for $\Bar{N}>1$); 
we do not report the finite temperature optimisation here, 
since it would require the unrealistic access to purifications of the bath \cite{genoni14}.
Notice also that, since the conditional covariance matrix (and hence the squeezing) after any general-dyne detection does
not depend on the measurement outcome, mere monitoring already achieves the optimal performance allowed by this class of 
feasible measurements, without the need of closing the control loop.\smallskip

\noindent{\em Simple Coherent Feedback --}
As a preliminary piece of inquiry, let us report the treatment \cite{serafini_quantum_2017}, and examine the performance of the simplest possible coherent feedback loop by feeding the output of one interface into the input of the other after undergoing losses. To do this we will consider a system mode coupled to two input fields, each through a Hamiltonian of the form given in (\ref{coupling}). To avoid ambiguity, we will use a subscript $e$ to refer to environmental white noise modes, and the the subscript $in$ when referring to the input interacting with the system through the Hamiltonian given in (\ref{coupling}). Adding coherent feedback involves setting $\hat{\boldsymbol{r}}_{in, 1} = \hat{\boldsymbol{r}}_{e, 1}$ and $\hat{\boldsymbol{r}}_{in, 2}(t) = \Phi(\hat{\boldsymbol{r}}_{out, 1}(t))$, where $\Phi$ is the CP-map corresponding to losses. These losses can be modelled as mixing at a beam splitter with an environmental mode $\hat{\boldsymbol{r}}_{e, 2}$. This means that coherent feedback can be achieved by setting :
\be \label{losses}
\begin{split}
    \rin2 & = \sqrt{\eta}\, \ro1  + \sqrt{1-\eta}\, \re2 \\ & = \sqrt{\eta}\, (\sqrt{\gamma} \rs - \re1)  + \sqrt{1-\eta}\, \re2 \, ,
\end{split}
\ee
where $\eta$ is the loss rate and we have used the input output relation (\ref{boundary}). It is important to note that we are assuming instantaneous  feedback, with no delays between the mode put out at interface 1 and fed back at interface 2, which will preserve the Markovianity of the dynamics. Making this substitution into Eq.~(\ref{coupling}) results in the system $\rs$ being effectively coupled to the environment $\re{tot} = (\re1^{\sf T}, \re2^{\sf T})^{\sf T}$ through the coupling matrix
$C = \sqrt{\gamma}
    (1- \sqrt{\eta}\, \Omega_1^{\sf T} , \sqrt{1-\eta}\, \Omega_1^{\sf T})$.
Such a system requires $\gamma(1-\sqrt{\eta})>\frac{\chi}{2}$ in order to be stable. The steady-state squeezing achieved in these conditions is $\sigma_{11} = \frac{\Bar{N}\gamma(1-\sqrt{\eta})}{\frac{\chi}{2} + \gamma (1-\sqrt{\eta})}$, which is minimised by letting $\sqrt{\eta} \rightarrow 1-\frac{\chi}{2 \gamma}$, 
resulting in a squeezing of $\sigma_{11} \rightarrow \frac{\Bar{N}}{2}$. 
Thus, at zero temperature (i.e., for $\Bar{N}=1$), {\em coherent feedback allows the 3dB limit to be approached (but not beaten)  
for any choice of parameters satisfying $0<\frac{\chi}{2\gamma} <1$}. Notice that, regardless of $\chi$, 
no stable squeezing is achievable if $\bar{N}\ge 2$.

This is a very remarkable result, showing that a coherent feedback loop is in principle capable to amplify the strength of any 
squeezing Hamiltonian up to the 3dB stability limit. 
However, a homodyne measurement-based loop, whose steady-state squeezing is given by Eq.~(\ref{moni}),
outperforms this coherent feedback scheme when the efficiency $\zeta$ of the detector satisfies 
\be
\zeta \ge \frac{2\left(\gamma-\chi\right)}{2\left(\gamma-\chi\right)+\Bar{N} \left(2\chi-\gamma\right) }  \label{condo}
\ee
and the denominator of the RHS of (\ref{condo}) is positive.
For $\chi< \gamma/2$, either the denominator is negative or the bound above is larger than $1$, which  
proves that homodyne monitoring does not beat coherent feedback at such weak interaction strengths. 
For $\chi\ge \gamma/2$, there is always a detection efficiency threshold above which the coherent feedback loop we considered 
is outperformed by homodyne monitoring; this threshold, quite interestingly, decreases with increasing noise (although 
the absolute performance of monitoring at given $\chi$ still deteriorates as the noise increases).
As the upper limit for stability $\chi=\gamma$ is approached, the efficiency threshold falls 
to zero, so that detection with any efficiency will be better than our coherent loop in this limit.
The ultimate performance of homodyne monitoring is obtained at $\zeta=1$, where $\sigma_{11}^m = \Bar{N}(1 - \chi/\gamma)$: 
hence, monitoring can in principle achieve stable squeezing ($\sigma_{11}^m<1$) for all values of $\Bar{N}$, although only for $\chi>\gamma(1-1/\Bar{N})$. 
Notice that, in principle, arbitrarily high squeezing may be stabilised at all noises (temperatures), whereas the coherent feedback 
loop we studied is bounded by the value $\Bar{N}/2$.
In order to achieve a conclusive comparison between coherent and measurement-based loops, 
we need to extend our treatment beyond a specific coherent feedback 
loop to include any possible interferometric scheme without additional sources of squeezing.\smallskip

\noindent{\em General Passive Coherent Feedback --}
Let us therefore consider the most general possible coherent feedback protocol which does not include any extra source of squeezing. A single system mode is coupled to $l+m$ input modes through a coupling Hamiltonian of the same form as (\ref{coupling}). 
We will give the label $a$ to the modes interacting at the first $l$ input-output interfaces, and the label $b$ to the modes 
interacting at the remaining $m$ interfaces. 
The first $l$ input modes are environmental white noise, meaning we can write $\rin a = \re a = (\re 1^{\sf T}...\re l
^{\sf T})^{\sf T}$. The corresponding output modes then undergo the most general Gaussian CP-map which does not include any form of squeezing. This is achieved by applying a passive transformation on the output modes, along with $n$ ancillary white noise modes before tracing out the ancillas. 
After the transformation, the resulting $m$ modes are then fed into the remaining $m$ input interfaces of the system. 
We shall assume that the additional ancillary modes are also affected by the same thermal noise
as the environment, so that it is still $\boldsymbol{\sigma}_{in}=\Bar{N}\id$
\footnote{Such modes clearly should not be squeezed, to avoid the unfair introduction of additional squeezing into the system; further, they should not be 
at a lower temperature than the original environment, since otherwise a hypothetical experimentalist could just replace the latter 
with one at a lower temperature, which would be a nice but highly questionable possibility in practice}.
Let us stress that, by considering the most general passive symplectic transformation, the formalism below 
will allow for an elegant and compact description of the most general interferometric scheme mediating the coherent feedback loop.

The passive transformation on the output and ancilla modes can be represented as an orthogonal, symplectic, $2(l+n)$-dimensional, square matrix $Z$:
\be
    \rin b \oplus \ranc f = Z (\ro a \oplus \ranc i ) \; ,
\ee
where $\rin b = (\rin{(l+1)}^{\sf T} \dots \rin{(l+m)}^{\sf T})^{\sf T}$ and $\ro a = (\ro1^{\sf T} \dots \ro{l}^{\sf T})^{\sf T}$. The initial and final states of the ancilla modes are given by $\ranc i$ and $\ranc f$ respectively where $\ranc i = (\re{(l+m+1)}^{\sf T}...\re{(l+m+n)}^{\sf T})^{\sf T}$

The orthogonal symplectic matrix $Z$ can be decomposed into block matrices $Z = \left( \begin{smallmatrix} 
        E & F \\
        G & H \\ 
\end{smallmatrix} \right)$.
This representation of $Z$ allows us to write $\rin b = E \ro a + F \ranc i$. It is shown in \cite{SM} that the overall effect of this coherent feedback protocol is to couple the system mode to the white-noise environment, now given by $\re a \oplus \ranc i$, through the coupling matrix:
\be
    C_{cf} = 
    \bmat
        C_l - C_m E\,\,&|\,\, C_m F \label{ccf}
    \emat \; ,
\ee
where $C_j$ indicates a $2 \x 2 j$ dimensional matrix of the form $\sqrt{\gamma}(\Omega^{\sf T} \dots \Omega^{\sf T})$. It is also shown that adding coherent feedback modifies the system Hamiltonian matrix $H_S$ by addition of a matrix: 
\be
H_{cf} = H_S+C_m E \Gamma_l + \Gamma_l^{\sf T} E^{\sf T} C_m^{\sf T} \; , \label{hcf}
\ee
where $\Gamma_l$ is a $2l\x2$ matrix of the form: $\Gamma_l = \sqrt{\gamma}(\id_2 \dots \id_2)^{\sf T}$.

\noindent{\em Optimal Coherent Feedback for Squeezing --}
Working within this general framework of coherent feedback, we will now find the optimal steady-state squeezing achievable. In particular, we will show that no coherent feedback protocol can improve upon the 3dB squeezing limit, for any choice of quadratic system Hamiltonian. 
This result will be outlined here, with some details left to \cite{SM}.

We are after the smallest eigenvalue of $\sig_{\infty}$, as given by Eq.~(\ref{steadystate}), with $D$ and $A$ which are modified by the 
coherent feedback loop as per Eqs.~(\ref{ccf},\ref{hcf}) (recall that $A$ and $D$ are in turn functions of $H_S$, $C$ and $\sig_{in}$),
which is equivalent to the smallest value of $\ve^{\dag} \sig_{\infty}\ve$ for a normalised vector $\ve$.
It may be shown that \cite{SM}, for any coherent feedback loop, the diffusion matrix is proportional to the identity and therefore has only a single eigenvalue given by 
$ \delta = \Bar{N}\gamma (l+m - 2 \epsilon)$ where $\epsilon = \sum_{jk}E^{jk}_{11}$ is the sum over the $11$-elements of each $2 \x 2$ submatrix $E^{jk}$ of $E$. 

This simplifies our task greatly, since it is now apparent that the smallest value of $\ve^{\dag} \sig_{\infty} \ve$ is obtained 
by setting $\ve = \boldsymbol{\lambda}_1$ where $\boldsymbol{\lambda}_1$ is the normalised eigenvector of $A$ corresponding to the  
eigenvalue $\lambda_1$ with most negative real part: indeed, this choice minimises the positive integrand $\ve^{\dag} {\rm e}^{A^{\sf T}t}
{\rm e}^{At} \ve$ {\em at all times} $t$. Whence the bound
\be
\ve^{\dag} \sig_{\infty}\ve \ge \delta \int_{0}^{\infty}  e^{(\lambda_1^* + \lambda_1) t} \,{\rm d}t 
= - \frac{\delta}{\lambda_1^* + \lambda_1} \, . \label{bound}
\ee
How negative the eigenvalue of $A$ can be made is limited by the stability criterion, since making one eigenvalue more negative makes the other less negative. To ensure stability, the most negative eigenvalue of $A$ must satisfy $\lambda_1 > \gamma(2 \epsilon -l-m)$ \cite{SM}. 
In deriving this bound, the system Hamiltonian was assumed to be quadratic, but otherwise completely general. This takes into account the modifications made to the system Hamiltonian due to the coherent feedback, as well as any deliberate tuning. Inequality (\ref{bound}) along with the expressions for $\delta$ and $\lambda_1$ yield
\be
    \ve^{\sf T} \sig_{\infty} \ve > \frac{\Bar{N}}{2} \; .
\ee
This result shows that there exists no combination of quadratic Hamiltonians and coherent feedback protocol which can beat the $3$dB limit.
Therefore, our comparison with the measurement-based strategy extends beyond the specific example we made above.
Summarising, we have found that, in regard to the squeezing Hamiltonian $-\frac{\chi}{4}\{\hat{x},\hat{p}\}$, coherent feedback loops are superior 
for $\chi<\gamma/2$ (with optimal performances independent from the interaction strength), while measurement-based, homodyne 
feedback is better for $\chi\ge \chi/2$ {\em and} efficiencies satisfying (\ref{condo}). 
Our comparison is definitive at zero temperature, for $\Bar{N}=1$, in the sense that both measurement-based and coherent feedback were fully optimised
for vacuum input noise (homodyning is then optimal), and that at optical frequencies one has $(\Bar{N}-1)\approx10^{-6}$.

It is also worth noting that our result hinges on the phase-insensitive nature of the input-output coupling (\ref{coupling}), which implies 
a diffusion matrix $D$ proportional to the identity, and would not apply, for instance, to a quantum Brownian motion master equation.
In this regard, our finding may be considered as an extension of the well-known 3dB squeezing limit that affects phase-insensitive amplifiers \cite{caves82}, which we showed to bound in-loop, {\em stable} squeezing too under any interferometric, coherent feedback scheme and any 
system quadratic Hamiltonian. Notice that stability is another essential ingredient in establishing the bound, as 
unstable coherent feedback loops would be able to achieve higher squeezing (but are typically not desirable in practice).\smallskip

\noindent{\em Conclusions and Summary --}
We have developed a general framework for passive coherent feedback in the Gaussian regime and shown 
that no protocol within this framework can beat the $3$dB squeezing limit at steady state. 
In contrast, homodyne monitoring of output fields can stabilise arbitrarily 
high squeezing at low enough noise and provided that detection efficiency is high enough. 

The general treatment developed here provides the groundwork for further inquiries on passive coherent feedback, which may be extended 
to the optimisation of entanglement in multimode systems, to more general noise models, as well as applied to the cooling of concrete systems, 
such as quantum optomechanics \cite{coherentoptomechanics_19}.\medskip

\acknowledgments
We acknowledge discussions with M. Brunelli, who made us aware of additional literature, and M. Genoni, 
who flagged inaccuracies in the manuscript.

\bibliography{CoherentFeedbackBibliography}

\newpage

\widetext

\begin{center}{\bf Supplemental Material}\end{center}

\section{Homodyne monitoring at finite temperature}

Continuous, general-dyne monitoring of the output field turns the diffusive equation (\ref{dde}) into the following Riccati equation (see, e.g., \cite{serafini_quantum_2017} for a complete treatment of the theory):
\be
\dot{\boldsymbol{\sigma}} = \tilde{A} \boldsymbol{\sigma} + \boldsymbol{\sigma} \tilde{A}^{\sf T} + \tilde{D} - 
 \boldsymbol{\sigma} BB^{\sf T}  \boldsymbol{\sigma} \; ,
\ee
for 
\begin{align}
\tilde{A} = A+GB^{\sf T} \; &, \quad \tilde{D} = D-GG^{\sf T}\; , \\
B= C \Omega ( \boldsymbol{\sigma}_{in}+ \boldsymbol{\sigma}_{m})^{-1/2} \; & \quad
G= \Omega C \boldsymbol{\sigma}_{in} ( \boldsymbol{\sigma}_{in}+ \boldsymbol{\sigma}_{m})^{-1/2} \, ,
\end{align}
where the covariance matrix $\boldsymbol{\sigma}_{m}$ parametrises the choice of measurement. We will consider the homodyne detection of the output field of a single-input single-output system as described in the section titled `\emph{Squeezing with no control}'.
Homodyne detection of the $\hat{x}$ quadrature with efficiency $\zeta$ is obtained by setting
\be
\boldsymbol{\sigma}_{m} = \lim_{z\rightarrow0} \begin{pmatrix} \frac{z+1-\zeta}{\zeta} & 0 \\
                              0 & \frac{\frac1z+1-\zeta}{\zeta}\end{pmatrix} \; ,
\ee
which leads to a diagonal quadratic equation for the monitored steady state covariance matrix, whose diagonal 
elements $\sigma^{m}_{11}$ and $\sigma_{22}^{m}$ 
must satisfy (the off-diagonal elements vanish)
\begin{align}
\frac{\gamma\zeta}{\zeta(\Bar{N}-1)+1} \sigma_{11}^{m\, 2} + \left(\gamma+\chi-2\frac{\gamma\zeta\Bar{N}}{\zeta(\Bar{N}-1)+1}\right)\sigma_{11}^{m} 
+ \frac{\gamma\zeta\Bar{N}^2}{\zeta(\Bar{N}-1)+1} -\gamma\Bar{N} =& 0 \, , \\
(\gamma-\chi)\sigma^{m}_{22} -\gamma \Bar{N} =& 0 \, ,
\end{align}  
with physical solutions
\begin{align}
 \sigma_{11}^m &= \frac{2 {\Bar{N}}\zeta -(1+(\Bar{N}-1)\zeta)(1+\frac{\chi}{\gamma})+\sqrt{[2 {\Bar{N}}\zeta -(1+(\Bar{N}-1)\zeta)(1+\frac{\chi}{\gamma})]^2+4 \Bar{N} \zeta (1-\zeta)}}{2\zeta}   \, ,  \\
\sigma_{22}^{m} &= \frac{\Bar{N} }{1-\frac{\chi}{\gamma}} \, . 
 \end{align}
The other solution for $\sigma_{11}^m$ must be discarded since it is negative or zero, and would thus violate the strict positivity of 
$\boldsymbol{\sigma}^{m}$, stemming from the uncertainty principle. 
The solution for $\sigma^{m}_{22}$ shows that, even under monitoring, the condition $|\chi|<\gamma$
is necessary for stability.

\section{Effective Coupling Matrix for General Passive Coherent Feedback}
The Hamiltonian which couples system and input modes can be written as:
\be
        \hat{H}_C = \frac{1}{2}\hat{\boldsymbol{r}}_{tot}^{\sf T} H_C \hat{\boldsymbol{r}}_{tot}  \; ,
\ee
where $\hat{\boldsymbol{r}}_{tot} = (\rs^{\sf T}, \rin a^{\sf T}, \rin b^{\sf T})^{\sf T}$. There are $l$ input modes at $a$ and $m$ input modes at $b$, so $\hat{\boldsymbol{r}}_{tot}$ is a $(2+2l+2m)$-dimensional vector. The coupling Hamiltonian matrix takes the form:
\be
    H_C = 
    \bmat
        0 & C_{l} & C_m \\
        C_l^{\sf T} & 0 & 0   \\
        C_m^{\sf T} & 0 & 0   \\
    \emat \; ,
\ee
where $C_j$ indicates a $2 \x 2j$ matrix of the form $\sqrt{\gamma}(\Omega^{\sf T} \dots \Omega^{\sf T})$. This allows for an exchange of excitations between  system and input field. When no coherent feedback is present, both $\rin a$ and $\rin b$ are white noise environmental modes. When coherent feedback is included, the input modes at $a$ are still white noise. We will call these modes $\re a$ to indicate this. The output modes at $a$ undergo a passive Gaussian CP-map and then are used to replace $\rin b$.

The passive Gaussian CP-map is achieved by performing a passive symplectic operation on the joint state $\ro a \oplus \ranc i$ where $\ranc i$ is a $2n$-dimensional vector representing the initial state of $n$ environmental white noise modes. The resulting mode to be input at interface $b$, along with the final state of the ancilla mode can be written $(\rin b \oplus \ranc f) = Z (\ro a \oplus \ranc i)$. The $2(l+n)$-dimensional square matrix $Z$ is symplectic, which ensures that the linear operation is physical, and orthogonal, which ensures that the operation is passive (i.e., that it does not involve any squeezing). We can write $Z$ in terms of block matrices:
\be
    Z =
    \bmat
        E & F\\
        G & H \\
    \emat \; .
\ee

Once the ancilla modes have been traced out, the effect of the CP-map can be written as $\ro a \rightarrow E \ro a + F \ranc i$ which 
allows us to write (note that $E$ and $F$ are, respectively, $2m\times 2l$ and $2m\times 2n$ matrices):
\be
    \rin b = 
    \bmat
        E & F \\
    \emat
    \bmat
        \ro a \\
        \ranc i \\
        \emat \; .
\ee

We now write the matrix form of the multimode input-output boundary condition in order to write $\ro a$ in terms of $\rin a = \re a$:
\be
    \bmat
        \ro a \\
        \ranc i \\ 
    \emat
    =
    \bmat
        \Gamma_l & -\id_l & 0 \\
        0      &   0  & \id_n\\
    \emat
    \bmat
        \rs \\
        \re a\\
        \ranc i \\
    \emat
    \quad \text{with} \quad
    \Gamma_l = \sqrt{\gamma}
    \bmat
        \id_2 \\
        \vdots \\
        \id_2 \\
    \emat \; .
\ee
Combining the above equations, we obtain:
\be
    \bmat
        \rs \\
        \rin a\\
        \rin b\\
    \emat
    =
    \bmat
        \id & 0 & 0 & 0 \\
        0 & \id & 0 & 0 \\
        0 & 0   & E & F \\
    \emat
    \bmat
        \id     & 0    & 0 \\
        0       & \id  & 0 \\
        \Gamma_l  & -\id  & 0 \\
        0       & 0    & \id
    \emat
    \bmat
        \rs \\
        \re a\\
        \ranc i \\
    \emat
    = L 
    \bmat
        \rs \\
        \re a\\
        \ranc i \\
    \emat \, .
\ee
The effect of adding coherent feedback is therefore to couple the system to a white noise environment given by $(\re a^{\sf T}, \ranc i^{\sf T})^{\sf T}$ through a coupling Hamiltonian characterised by the matrix $H_C^{cf} = L^{\sf T} H_C L$. This matrix is:
\be
    H_C^{cf} =
    \bmat
        C_m E \Gamma_l + \Gamma_l^{\sf T} E^{\sf T} C_m^{\sf T} & C_l - C_m E & C_m F\\
        C_l^{\sf T} -E^{\sf T}C_m^{\sf T} & 0 & 0 \\
        F^{\sf T}C_m^{\sf T} & 0 & 0 \\
    \emat \; ,
\ee
which couples the system to the environment through the Hamiltonian operator
\be
    \hat{H}_C^{cf} = \frac{1}{2} (\rs^{\sf T}, \re a^{\sf T}, \ranc i ^{\sf T} ) H_C^{cf} (\rs, \re a, \ranc i  ) \; .
\ee
Notice that this results in a matrix equal to $C_m E \Gamma_l + \Gamma_l^{\sf T} E^{\sf T} C_m^{\sf T}$ being added to the system Hamiltonian matrix and changes the effective coupling matrix to:
\be \label{CFcouplingmatrix}
    C_{cf} = (C_l - C_m E \quad | \quad C_mF) \; .
\ee

\section{Properties of the Orthogonal Symplectic Matrix}

We have considered an orthogonal symplectic matrix of the form
\be Z =
    \bmat
    E & F \\
    G & H \\
    \emat \; , \label{zed} 
\ee
which transformed a vector of operators as per $\hat{\boldsymbol{r}} \mapsto{} Z \hat{\boldsymbol{r}} $. Here, $E$ is a $(2m\x2l)$ matrix and $F$ is a $(2m\x 2n)$. The condition of orthogonality means that $ZZ^{\sf T} = \id$, which gives us the following conditions on the submatrices:

\be
    ZZ^{\sf T} =
    \bmat
        E & F\\
        G & H \\
    \emat
    \bmat
        E^{\sf T} & G^{\sf T}\\
        F^{\sf T} & H^{\sf T} \\
    \emat
    =
    \bmat
        EE^{\sf T} + FF^{\sf T} & EG^{\sf T} + F H^{\sf T}\\
        G E^{\sf T} + H F^{\sf T} & GG^{\sf T} + HH^{\sf T} \\
    \emat
    =
    \bmat
        \id & 0 \\
        0 & \id \\
    \emat
\ee
In particular, we shall make use of the relation $EE^{\sf T} +FF^{\sf T} = \id$. The condition of symplecticity means that $Z\Omega Z^{\sf T} = \Omega$. Recall that we are using the convention that the dimension of $\Omega$ is specified by the context. In terms of the submatrices, this means that:
\be
    Z\Omega Z^{\sf T} =
    \bmat
        E & F\\
        G & H \\
    \emat
    \bmat
        \Omega & 0 \\
        0 & \Omega \\
    \emat
    \bmat
        E^{\sf T} & G^{\sf T}\\
        F^{\sf T} & H^{\sf T} \\
    \emat
    =
    \bmat
        E\Omega E^{\sf T} + F\Omega F^{\sf T} & E\Omega G^{\sf T} + F \Omega H^{\sf T}\\
        G\Omega E^{\sf T} + H\Omega F^{\sf T} & G\Omega G^{\sf T} + H\Omega H^{\sf T} \\
    \emat
    =
    \bmat
        \Omega & 0 \\
        0 & \Omega \\
    \emat \, .
\ee
From this we obtain the condition $E \Omega E^{\sf T} + F \Omega F^{\sf T} =\Omega$, which will be key later.

The vector of operators $\hat{\boldsymbol{r}}$ was ordered so that $\hat{\boldsymbol{r}} = (\hat{x}_1, \hat{p}_1 ... \hat{x}_n, \hat{p}_n)^{\sf T}$. We can also consider an orthogonal symplectic matrix $S$ acting on a vector of differently ordered operators: $\hat{\boldsymbol{s}} \xrightarrow{} S \hat{\boldsymbol{s}}$ where $\hat{\boldsymbol{s}} = (\hat{x}_1 ... \hat{x}_n, \hat{p}_1 ... \hat{p}_n)$. In this case, the transformation matrix takes the form \cite{serafini_quantum_2017}:
\be
    S =
    \bmat
        X & Y \\
        -Y & X \\
    \emat \quad \text{with} \quad XY^{\sf T} - YX^{\sf T} = 0_n \quad \text{and} \quad XX^{\sf T} + YY^{\sf T} = \id_n \; .
\ee
When we use the ordering of variables $\hat{s} = (\hat{x}_1 \dots \hat{x}_n, \hat{p}_1\dots \hat{p}_n)^{\sf T}$, the symplectic condition 
is $SJS^{\sf T} = J$, where $J$ is the symplectic form
\be
    J = 
    \bmat
        0_n   & \id_n \\
        -\id_n & 0_n   \\
    \emat \; .
\ee
Transforming between the two representations means that we can write each $2 \x 2$ submatrix of $Z$ as:

\be \label{submatrices}
    Z = 
    \bmat
        Z_{11} & \dots & Z_{1n} \\
        \vdots & \ddots & \vdots \\
        Z_{n1} & \dots & Z_{nn} \\
    \emat
    \quad Z_{jk} = 
    \bmat
        x_{jk} & y_{jk} \\
        -y_{jk} & x_{jk} \\
    \emat \; ,
\ee
where $x_{jk}$ and $y_{jk}$ are the elements of matrices $X$ and $Y$ respectively. This fact will be used later.

\section{Eigenvalues of the Drift Matrix}
The drift matrix $A$ can be expressed in terms of the Hamiltonian and coupling matrices $H_S$ and $C$ as
\be
    A = \Omega H_S + \frac{1}{2}\Omega C \Omega C^{\sf T} \; .
\ee
Note that $C \Omega C^{\sf T}$ is an $2 \x 2$ antisymmetric matrix, since $\Omega^{\sf T} = - \Omega$. Therefore, for a single mode, $\frac{1}{2}\Omega C \Omega C^{\sf T}$ is proportional to the identity. We shall set $\frac{1}{2}\Omega C \Omega C^{\sf T} = \beta \id$. Also, since $H_S$ is a symmetric matrix, $Tr[\Omega H_S]=0$, meaning that the eigenvalues of $\Omega H_S$ can be written as $\pm h$ and the eigenvalues of $A$ can be written $\lambda = \beta \pm h$. 

We will now find the value of $\beta$ in the coherent feedback framework, where Eq.~(\ref{CFcouplingmatrix}) determines the coupling matrix. 
We use the notation $\Gamma_k$ from earlier to indicate a $2k \x 2$-dimensional matrix of the form $\Gamma_k= \sqrt{\gamma}(\id \dots \id)^{\sf T}$. 
This satisfies $\Omega C_k= \Gamma_k^{\sf T}$ and $\Omega C_k ^{\sf T} = - \Gamma_k $. We can write
\be
\begin{split}{}
    \Omega C_{cf} \Omega C_{cf}^{\sf T} & = (\Omega C_l - \Omega C_mE)(\Omega C_l^{\sf T} - \Omega E^{\sf T} C_m^{\sf T}) + \Omega C_m F \Omega F^{\sf T} C_m^{\sf T} \\ & 
    =  -\Gamma_l^{\sf T} \Gamma_l - \Gamma_l^{\sf T} \Omega E^{\sf T} C_m^{\sf T} + \Gamma_m^{\sf T}E\Gamma_l + \Gamma_m^{\sf T} E \Omega E^{\sf T} C_m^{\sf T} +\Gamma_m^{\sf T} F \Omega F^{\sf T} C_m^{\sf T} \, .
\end{split} 
\ee
We now use the symplectic property $E\Omega E^{\sf T} + F \Omega F^{\sf T} = \Omega$, derived in the previous section, to write 
\be
    \Omega C_{cf} \Omega C_{cf}^{\sf T} = -\Gamma_l^{\sf T} \Gamma_l 
    - \Gamma_l^{\sf T} \Omega E^{\sf T} C_m^{\sf T} 
    + \Gamma_m^{\sf T}E\Gamma_l + \Gamma_m^{\sf T}  \Omega C_m^{\sf T} \, 
\ee
(recall that the notation $\Omega$ refers to symplectic forms of different dimension, as appropriate for matrix multiplications to be consistent).
Noting now that $\Gamma_k^{\sf T}\Gamma_k = k\gamma \id$, one has
\be
    \Omega C_{cf} \Omega C_{cf}^{\sf T} = -(l+m)\gamma \id - \Gamma_l^{\sf T} \Omega E^{\sf T} C_m^{\sf T} + \Gamma_m^{\sf T} E \Gamma_l \; ,
\ee
which can be written in terms of the $2 \x 2$ submatrices of $E$:
\be
    E = 
    \bmat
        E_{11} & \dots & E_{1 l} \\
        \vdots & \ddots & \vdots \\
        E_{m1} & \dots & E_{ml} \\
    \emat
    \quad \quad
    E_{jk} = 
    \bmat
        e_{11}^{jk} & e_{12}^{jk} \\
        e_{21}^{jk} & e_{22}^{jk}\\
    \emat \; .
\ee
The matrix $\Gamma_m^{\sf T} E \Gamma_l$ can be written as $\gamma \sum_{i,j}E_{jk}$, while
$\Gamma_l^{\sf T} \Omega E^{\sf T} C_m^{\sf T}$ can be calculated in the same way:
\be
\begin{split}
    \Gamma_l^{\sf T} \Omega E^{\sf T} C_m^{\sf T} &  = \sqrt{\gamma} \Gamma_l^{\sf T} \Omega 
    \bmat
        \sum_{j=1}^m E_{j1}^{\sf T} \Omega \\
        \vdots \\
        \sum_{j=1}^m E_{jl}^{\sf T} \Omega
    \emat
    = \sqrt{\gamma} \Gamma_l^{\sf T}
    \bmat
        \sum_{j=1}^m \Omega E_{j1}^{\sf T} \Omega \\
        \vdots \\
        \sum_{j=1}^m \Omega E_{jl}^{\sf T} \Omega
    \emat
    = \gamma \sum_{j, k} \Omega E^{\sf T}_{jk} \Omega 
      \\ &= \gamma \sum_{j, k}
    \bmat
        0 & 1 \\
        -1 & 0\\
    \emat
    \bmat
        e_{11}^{jk} & e_{21}^{jk} \\
        e_{12}^{jk} & e_{22}^{jk}\\
    \emat
    \bmat
        0 & 1 \\
        -1 & 0\\
    \emat =  \gamma \sum_{j, k} 
    \bmat
        -e_{22}^{jk} & e_{12}^{jk}\\
        e_{21}^{jk} & -e_{11}^{jk}\\
    \emat \, .
\end{split}
\ee
Now, we use Eq.~(\ref{submatrices}) to write $e^{jk}_{11} = e^{jk}_{22}$. This allows us to put together the above results to obtain $\beta = \frac{\gamma}{2} (2 \epsilon- l - m)$ where $\epsilon = \sum_{j,k}e^{jk}_{11} = \sum_{j,k}e^{jk}_{22}$ [since $e^{jk}_{11}=e^{jk}_{22}$ as in Eq.~(submatrices)]. 

Recall that the two eigenvalues of $A$ are $\lambda = \beta \pm h$. In order for the system to be stable, we must have $Re[\lambda]<0$ for both eigenvalues. This requires that $\beta$ is negative. It also means that the most negative eigenvalue of $A$ cannot be lower than $2\beta$, since this would mean that the other eigenvalue would violate the stability criterion. We have therefore obtained the bound $\lambda_1 > \gamma(2\epsilon -l -m)$ on the most negative eigenvalue of $A$.

\section{Eigenvalues of the Diffusion Matrix}
The diffusion matrix $D$ takes the form $D = \Omega C \sig_{in}C^{\sf T} \Omega^{\sf T}$. Notice that since $\Omega$ is a unitary matrix, the eigenvalues of $D$ are the same as the eigenvalues of $C \sig_{in} C^{\sf T}$. The input state is taken to be a vacuum or thermal state 
with uniform noise, so $\sig_{in} = \Bar{N}\id$ with $\Bar{N}\geq 1$. This means that we can find the eigenvalues of $D$ for coherent feedback by finding the eigenvalues of the matrix $C_{cf}C_{cf}^{\sf T}$ and multiplying them by $\Bar{N}$:
\be
    C_{cf}C_{cf}^{\sf T} = C_lC_l^{\sf T} - C_mEC_l^{\sf T} - C_l E^{\sf T} C_m^{\sf T} + C_mEE^{\sf T} C_m^{\sf T} + C_mFF^{\sf T}C_m^{\sf T} \; .
\ee
Using the orthogonality condition $EE^{\sf T} + FF^{\sf T} = \id$ and the fact that $C_k C_k^{\sf T} = k \gamma \id_2$, we obtain:
\be
    C_{cf}C_{cf}^{\sf T} = \gamma(l+m)\id - C_mEC_l^{\sf T} - C_lE^{\sf T}C_m^{\sf T}\; .
\ee
Writing in terms of the $2\x2$ submatrices of $E$ gives:
\be
    C_{cf}C_{cf}^{\sf T} = \gamma(l+m)\id + \gamma \sum_{j,k} 
    \bmat
        -2 e_{22}^{jk} & e_{12}^{jk} + e_{21}^{jk} \\
        e_{12}^{jk} + e_{21}^{jk} & -2 e_{11}^{jk} 
    \emat
    = \gamma(l+m)\id -2\gamma \epsilon \id \; ,
\ee
where we have used $e^{jk}_{12} = -e^{jk}_{21}$ and $e^{jk}_{11} = e^{jk}_{22}$, as per Eq.~(\ref{submatrices}), and $ \epsilon = \sum_{jk}e^{jk}_{11}=\sum_{jk}e^{jk}_{22}$. Therefore, the diffusion matrix under coherent feedback is proportional to the identity with 
eigenvalue $\delta = \Bar{N} \gamma(l+m - 2 \epsilon)$.

\end{document}